\documentstyle[11pt,epsfig,amsfonts]{article}
\begin{document}
\title{Degenerate BPS Domain Walls: \\ Classical and Quantum Dynamics}
\author{A. Alonso Izquierdo$^{(a)}$,
M.A. Gonzalez Leon$^{(a)}$ \\ W. Garcia Fuertes$^{(b)}$,  M. de la
Torre Mayado$^{(c)}$ and J. Mateos Guilarte$^{(c)}$
\\ {\normalsize {\it $^{(a)}$ Departamento de Matematica
Aplicada}, {\it Universidad de Salamanca, SPAIN}}\\{\normalsize
{\it $^{(b)}$ Departamento de Fisica}, {\it Universidad de Oviedo,
SPAIN}} \\ {\normalsize {\it $^{(c)}$ Departamento de Fisica,}
{\it Universidad de Salamanca, SPAIN}}}

\date{}
\maketitle
\begin{abstract}
We discuss classical and quantum aspects of the dynamics of a
family of domain walls arising in a generalized Wess-Zumino model.
These domain walls can be embedded in ${\cal N}=1$ supergravity as
exact solutions and are composed of two basic lumps.
\end{abstract}
\section{Introduction}
Currently the topic of supersymmetric extended objects is
extremely fashionable. Before the advent of the new brane world,
however, only a few workers paid attention to the physical and
mathematical properties of super-membranes of various dimensions.
Among such pioneers, we mention the work on the cohomological
interpretation of the topological charges associated with these
extended objects by J. A. de Azcarraga and collaborators, see
\cite{Azca}. Relevant contributions to the subject can be found
also in the work of M. Cvetic, S. J. Rey et al, \cite{Rey}. In
this paper, we offer a brief summary of our work on a related
topic -the dynamics of BPS domain walls- to honor Adolfo,
Professor and friend to several of us from the Salamanca years
circa 1977.

We focus on a generalized Wess-Zumino model with two ${\cal N}=1$
chiral superfields, first discussed by Bazeia et al. in Reference
\cite{B1}. Slightly later in \cite{Voloshin}, it was shown by
Shifman and Voloshin  that this model admits a degenerate family
of BPS domain walls. The general variety of both non-BPS and BPS
solitary waves has been described in \cite{Aai3}, studying the
(1+1)-dimensional version of the system. More recently, Eto and
Sakai, see \cite{Eto}, have discovered how to define a \lq\lq
local" superpotential in such a way that the domain walls of the
generalized Wess-Zumino model remain exact solutions in ${\cal
N}=1$ (3+1)-dimensional supergravity.

A remarkable feature of this supersymmetric system is the
availability of analytic descriptions of the domain wall dynamics
along orthogonal lines to the \lq\lq two"-branes. BPS wall/BPS
anti-wall dynamics have been discussed in \cite{Sakai}, analyzing
the energy density of non-BPS wall/anti-wall configurations. In
\cite{Aai1}, however, several of us unveiled the adiabatic
dynamics of BPS two-walls by studying geodesic motion in the
moduli space. The dynamics inside the wall at low energy is ruled
by the \lq\lq effective action", see \cite{Gomis}, governing the
evolution of Goldstone bosons through the two-brane. Although
Lorentz invariance forbids dependence on the center of mass of the
wall, in our system with two real scalar fields the effective
action depends on the relative coordinate that labels the distance
between walls; the inertia for Goldstone bosons running either on
distant or intersecting walls are different, smoothly varying from
one to another.

The above results concerning the classical dynamics of domain
walls are based on a crucial pro\-perty: the degeneracy of the
classical moduli space of domain walls. The question arises as to
whether this degeneracy survives quantum fluctuations. Analyse of
the one-loop fluctuations around the wall solutions in the \lq\lq
body" of the supersymmetric system reveal that repulsive forces,
decaying exponentially with distance, arise between the
fundamental lumps, see \cite{Aai5}. However, a general theorem
warranting the identity between the one-loop corrections to kink
masses and the anomaly in the central charge of the ${\cal N}=1$
SUSY algebra, see e.g. \cite{Fuj}, tells us that at the quantum
level wall degeneracy occurs in the fully supersymmetric system.

\section{Moduli space of solitary waves in generalized \\ Wess-Zumino
models}

We shall consider the (3+1)-dimensional ${\cal N}=1$
supersymmetric Wess-Zumino model, where the superpotential
\[
W^{3D}(\Phi_1,\Phi_2)={4\over 3}\Phi_1^3-\Phi_1+2\sigma
\Phi_1\Phi_2^2
\]
determines the interactions between the two chiral
superfields\footnote{We shall use non-dimensional field variables
and coupling constants throughout the paper in order to keep the
formulas simpler.}: $\Phi_1(x^0,\vec{x},\theta_\alpha),\ \
\Phi_2(x^0,\vec{x},\theta_\alpha)$ . Here, the four-vectors
$x^\mu=(x^0,\vec{x})$ and the Grassman Weyl spinors
$\theta_\alpha$ provide (non-dimensional) coordinates in ${\cal
N}=1$ Minkowski superspace. $\sigma$ is the only (non-dimensional)
coupling constant.

In our search for domain walls, we need to explore only the \lq\lq
body" of the theory, i.e. we shall focus on the first terms of the
Grassman expansion of the fields and the superpotential:
$\Phi_1|_{\theta_\alpha=0}=\phi_1+i\psi_1,\ \
\Phi_2|_{\theta_\alpha=0}=\phi_2+i\psi_2$. Moreover, the reality
condition $\psi_1=\psi_2=0$ and the requirement of independence of
the $(y,z)$ variables (dimensional reduction),
$\phi_1(x^0,\vec{x})=\phi_1(x^0,x),\ \
\phi_2(x^0,\vec{x})=\phi_2(x^0,x) $, lead us to the
(1+1)-dimensional superpotential:
\[
W(\phi_1,\phi_2)={1\over 2}{\rm
Re}W^{3D}(\phi_1(x^0,x),\phi_2(x^0,x))={1\over 2}\left[ {4\over
3}\phi_1^3-\phi_1+2\sigma\phi_1\phi_2^2 \right].
\]
Therefore, the domain walls of the original Wess-Zumino model are
in one-to-one correspondence with the solitary waves (kinks) of
the (1+1)-dimensional system, with dynamics governed by the
action:
\begin{equation}
S=\int \, d^2 x\left[ {1\over
2}\partial_\mu\phi_1\partial^\mu\phi_1+{1\over
2}\partial_\mu\phi_2\partial^\mu\phi_2-{1\over 2}{\partial W\over
\partial \phi_1}{\partial W\over
\partial \phi_1} -{1\over 2}{\partial W\over
\partial \phi_2}{\partial W\over
\partial \phi_2}  \right]\ .\label{eq:bnrt}
\end{equation}
The vacuum moduli space, characterized as the set of critical
points of $W$ modulo the internal parity symmetry group of the
problem, contains the \lq\lq two" points: $
(\phi_1^{V_1^\pm}=\pm{1\over 2} \ , \ \phi_2^{V_1^\pm}=0) $ , $
(\phi_1^{V_2^\pm}=0 \ , \
\phi_2^{V_2^\pm}=\pm{1\over\sqrt{2\sigma}})$.
\subsection{The search for Kinks}
We shall focus only on the topological sector connecting the
$V_1^\pm$ vacua. Generically, solitary waves in other topological
sectors are not BPS kinks; the problem of kink stability is
studied in \cite{Aai4} from a geometrical point of view. The
energy for static configurations can be written \`{a} la
Bogomolny, resulting in:
\begin{equation}
E (\phi_1,\phi_2)={1\over 2}\int \, dx \left[ \left( {d\phi_1\over
dx}-{\partial W\over
\partial\phi_1}\right)^2+ \left({d\phi_2\over dx}-{\partial W\over
\partial\phi_2}\right)^2\right]+\int_P \, \left[ d\phi_1{\partial
W\over\partial\phi_1}+d\phi_2{\partial
W\over\partial\phi_2}\right].
\end{equation}
Given a polynomial superpotential such as $W$, the solutions of
the first-order equations
\begin{eqnarray} {d\phi_1\over dx}={\partial W\over\partial\phi_1} \qquad
,\qquad {d\phi_2\over dx}={\partial W\over\partial\phi_2}
\label{eq:flow} \end{eqnarray} are absolute minima of $E$, usually
referred to as BPS kinks that saturate the topological bound:
$E_T=\int_P \, dW=|W(+\infty)-W(-\infty)|$ .

From (\ref{eq:flow}), the flow lines of ${\rm grad}W$ are
identified as the solutions of the ODE:
\[
{d\phi_1\over d\phi_2}={4\phi_1^2+2\sigma\phi_2^2-1\over
4\sigma\phi_1\phi_2}\ .
\]
There is an integrating factor, $|\phi_2|^{-({2\over\sigma}+1)}$
if $\sigma\neq 0$, $\sigma\neq 1$, and the flow lines -Kink
orbits- are the curves:
\begin{equation}
\phi_1^2+{\sigma\over 2(1-\sigma)}\phi_2^2={1\over
4}+{c\over2\sigma}|\phi_2|^{{2\over\sigma}} \qquad ,
\label{eq:tra1}
\end{equation}
where $c \in (-\infty,c^S=\frac{1}{4} \frac{\sigma}{1-\sigma}
\left( 2\sigma\right)^{\frac{\sigma+1}{\sigma}})$ is an
integration constant.

The meaning of these solutions can be summarized as follows: there
are two maxima of $-U(\phi_1,\phi_2)$ with the same height. Kink
solutions which pass from one maximum to the other depend on a
parameter, $c$, which measures whether the particle moves through
the bottom of the valley or more along the sides on the curve
(\ref{eq:tra1}). There is a critical value $c^S$ of $c$ where the
particle moves as high as possible; when $c$ increases beyond this
critical value the particle crosses the mountain and falls off to
the other side, see Figure 1.

\begin{figure}[htbp]
\centerline{\epsfig{file=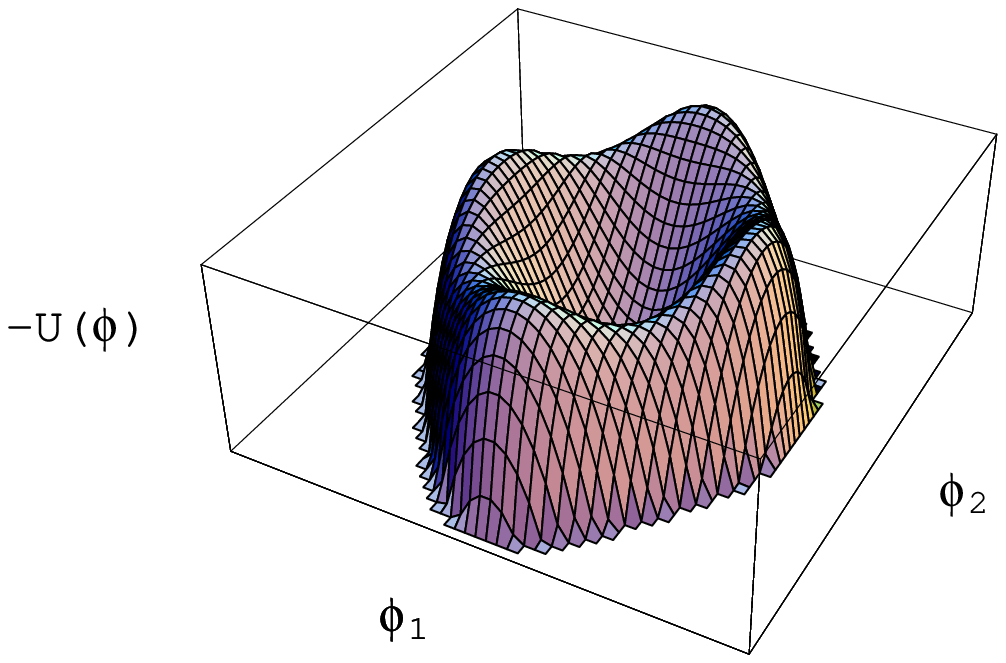, height=3.5cm}
\hspace{1cm}\epsfig{file=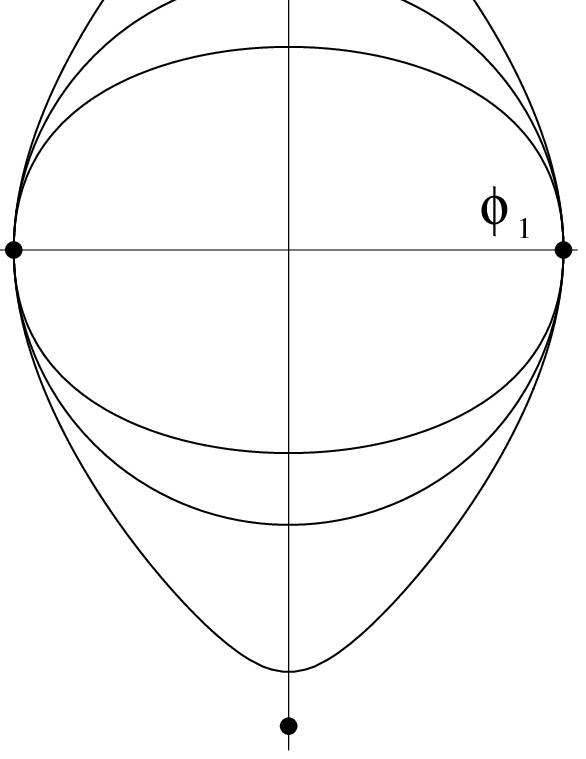, height=3.5cm}
\hspace{1cm} \epsfig{file=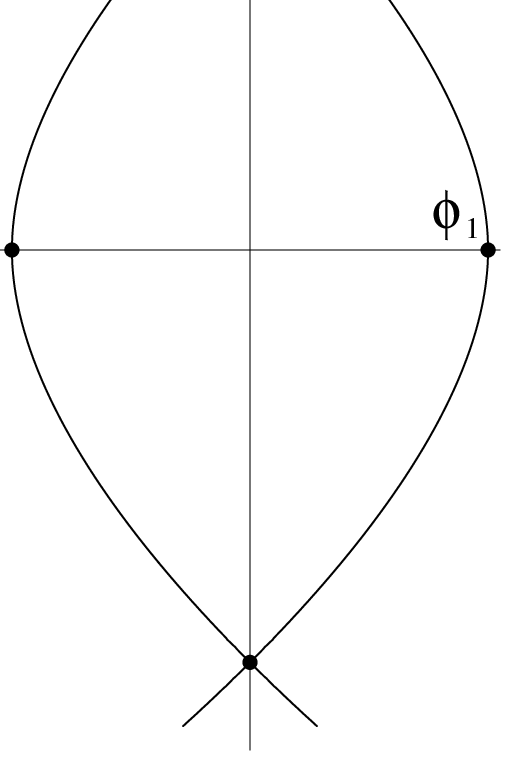,height=3.5cm}
\hspace{1cm}\epsfig{file=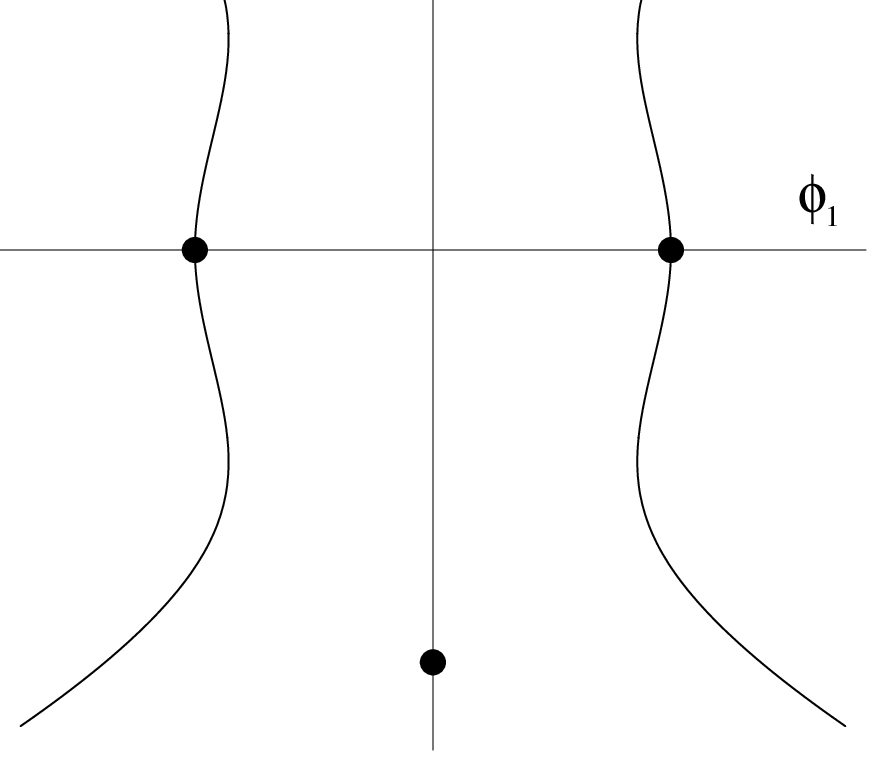, height=3.5cm}}
\caption{\small \it The $-U(\phi)=-{1\over 2}{\partial W\over
\partial \phi_1}{\partial W\over
\partial \phi_1} -{1\over 2}{\partial W\over
\partial \phi_2}{\partial W\over
\partial \phi_2}$ potential (left) Flow-lines: in the ranges $c
\in (-\infty,c^S)$ (middle left), $c=c^S$ (middle right), and $c
\in (c^S ,\infty)$ (right).}
\end{figure}
Exactly at the critical value, the kink orbit starts at the point
$\vec{\phi}^{V_1^\pm}$ and ends at the other point
$\vec{\phi}^{V_2^\pm}$; this is in contrast to any other kink
orbit for $c<c^S$, which starts at $\vec{\phi}^{V_1^\pm}$ but ends
in $\vec{\phi}^{V_1^\mp}$. Thus, there are two kinds of kinks
living in different topological sectors of the system: \lq\lq
link" kinks, interpolating between different points of the vacuum
moduli space, and \lq\lq loop" kinks, joining vacua identified as
the same point of the vacuum moduli.

To find the kink form factors, one plugs formula (\ref{eq:tra1})
into (\ref{eq:flow}) so that the problem is reduced to solving the
quadrature:
\begin{equation}
I[\phi_2]=\int \frac{d \phi_2}{\phi_2 \sqrt{\frac{1}{4}+\frac{c}{2
\sigma} |\phi_2|^{\frac{2}{\sigma}}-\frac{\sigma}{2(1-\sigma)}
\phi_2^2}}=2\sigma(x+a)=z\label{eq:quad}
\end{equation}
\[
\phi_1^{{\rm TK}2}[x;a,c]= \pm \sqrt{\frac{1}{4}+\frac{c}{2
\sigma} |I^{-1}(z)|^{\frac{2}{\sigma}}-\frac{\sigma}{2(1-\sigma)}
[I^{-1}(z)]^2}\quad , \quad \phi_2^{{\rm TK}2}[x;a,c]=I^{-1}(z)
\]
\subsection{Special cases: Liouville systems}
Explicit analytic integration of (\ref{eq:quad}) in terms of
elementary functions is only possible if $\sigma=2$ and $
\sigma={1\over 2}$. The reason is that the analogous mechanical
problem that one needs to solve in the search for one-dimensional
solitary waves is an integrable Liouville system. Also, when
$\sigma=3, 4, {1\over 3}$ and ${1\over 4}$, the quadrature can be
found analytically, but in these cases one is forced to deal with
elliptic functions.

We present the analytic outcome of finding $I$ and its inverse
$I^{-1}$ in the two Liouville cases. In Figure 2 we show kink
profiles for several values of $b$ and $\sigma={1\over 2}$.

$\bullet$   ${\bf\sigma}=2$
\[
\phi_1^{{\rm TK2}}[x;a,b]=\frac{(-1)^\alpha}{2} \frac{\sinh 4
(x+a)}{\cosh 4(x+a)+b} \hspace{1cm} \phi_2^{{\rm TK2}}[x;a,b]=
\frac{(-1)^\beta}{2} \frac{\sqrt{b^2-1}}{\cosh 4(x+a)+b}\qquad ,
\]
where $\alpha ,\beta=0,1$, and $a\in {\Bbb R}$ is the center of
the kink. The parameter $b$ is related to the integration constant
as follows: $b=\frac{-c}{\sqrt{c^2-16}}$, so that
$b\in(1,\infty)$.

$\bullet$  ${\bf\sigma}=\frac{1}{2}$
\begin{equation}
\phi_1^{{\rm TK2}}[x;a,b]= \frac{(-1)^\alpha}{2} \frac{\sinh
(x+a)}{\cosh (x+a)+b^2} \hspace{1cm} \phi^{{\rm
TK2}}_2[x;a,b]=(-1)^\beta\frac{b}{\sqrt{b^2+ \cosh
(x+a)}}\label{eq:2wal}\quad .
\end{equation}
Again, $a\in {\Bbb R}$ is the kink center, $b$ is related to $c$
as $b^2=\frac{1}{\sqrt{1-4 c}}\in (0,\infty)$, $\alpha
,\beta=0,1$.

\noindent
\begin{figure}[htbp]
\centerline{ \epsfig{file=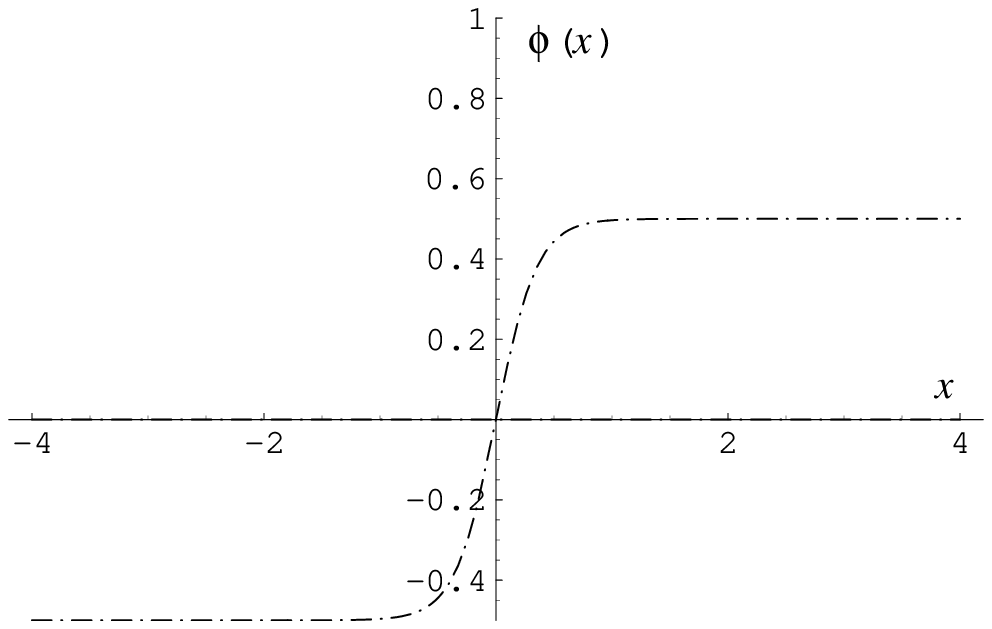, height=2.5cm}
\epsfig{file=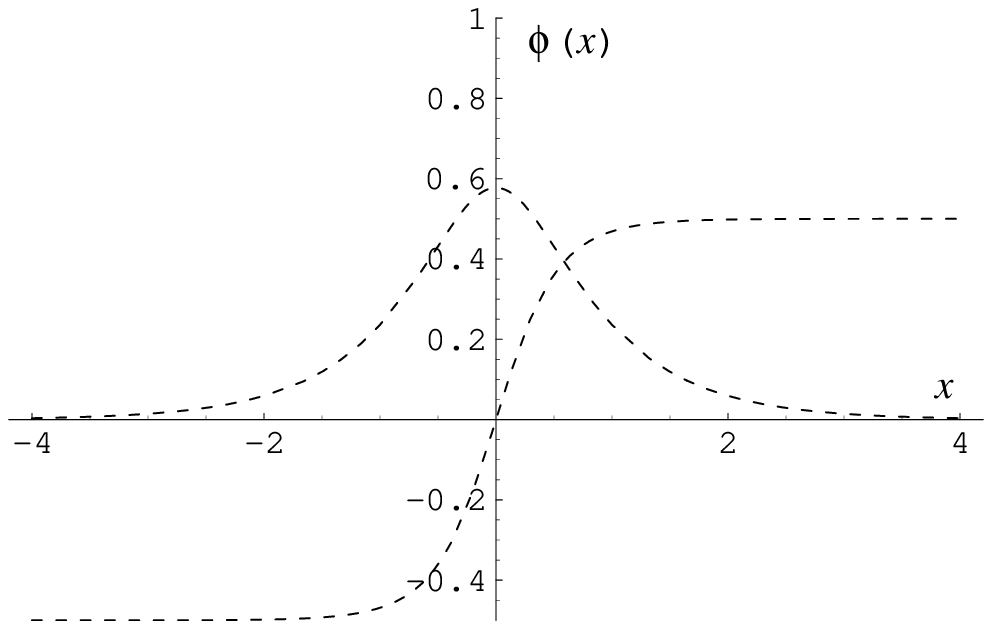, height=2.5cm}
\epsfig{file=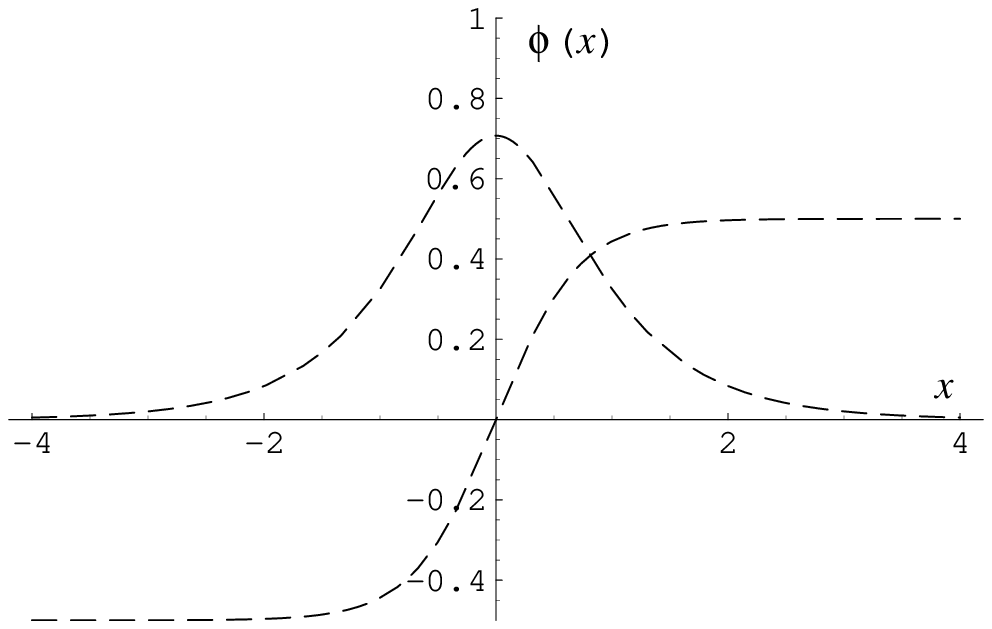, height=2.5cm}
\epsfig{file=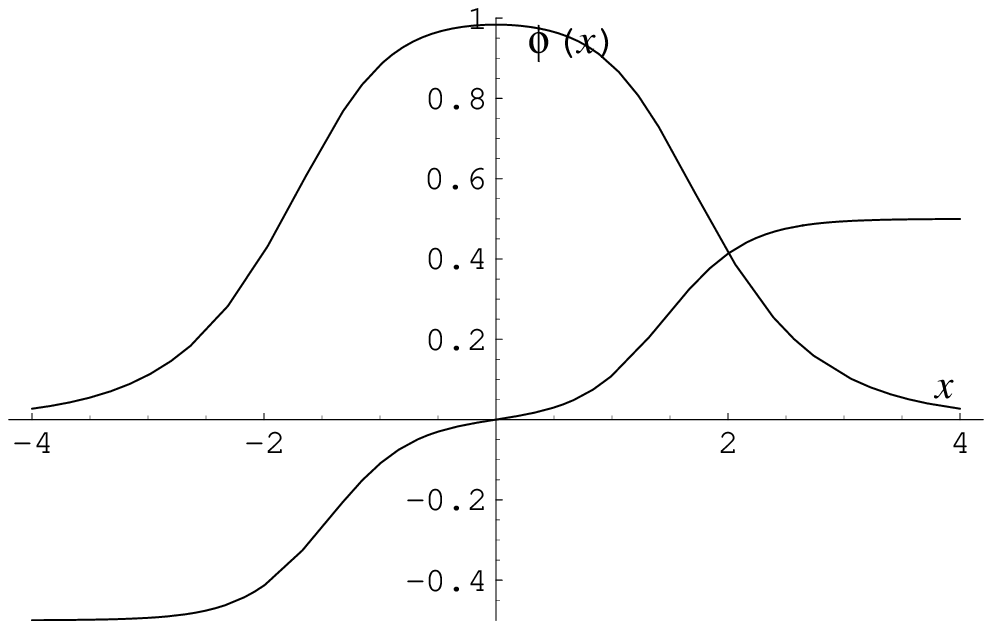, height=2.5cm}} \caption{\small
Solitary waves for $\sigma={1\over 2}$ corresponding to: {\it
(a)} $b=0$, {\it (b)} $b=\sqrt{0.5}$, {\it (c)} $b=1$ and {\it
(d)} $b=\sqrt{30}$.}
\end{figure}

\subsection{Moduli space of BPS kinks}
To elucidate the physical meaning of the $b$ parameter, we focus
on the $\sigma=\frac{1}{2}$ case because it provides an analytical
description of the generic behaviour. In Figure 3, pictures of the
energy density are depicted for the same kinks shown in Figure 2.
Note that for $b^2>1$ the energy density presents two lumps,
whereas if $b^2<1$ the density is of the usual bell-shaped form.
Also, because changing $b$ to $-b$ in the solution is equivalent
to changing $\phi_2$ to $-\phi_2$ and the energy density is not
sensitive to the sign of $b$, it is sensible to describe the
moduli space of kinks as the half-plane parametrized by the
$(a,b^2)$ coordinates: $a$ fixes the center of mass of the two
lumps, and $b^2$ can be interpreted as the relative coordinate
that measures the distance between them.
\begin{figure}[htbp]
\centerline{ \epsfig{file=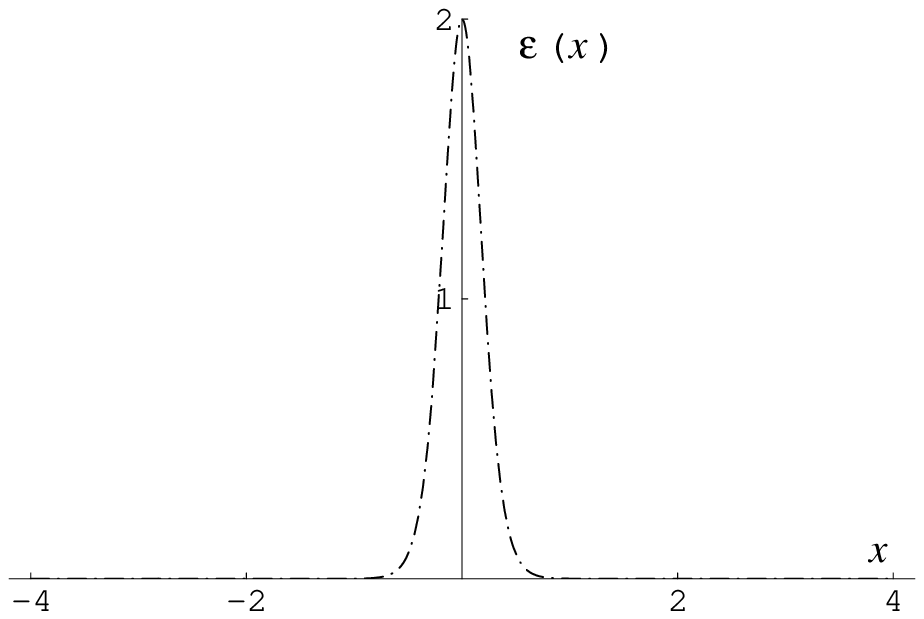, height=2.5cm}
\epsfig{file=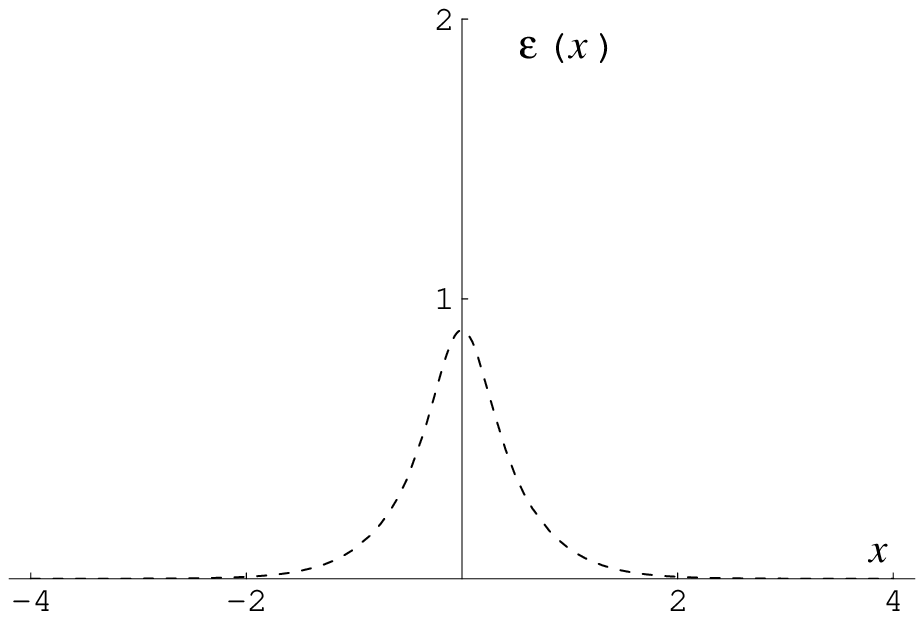, height=2.5cm}
\epsfig{file=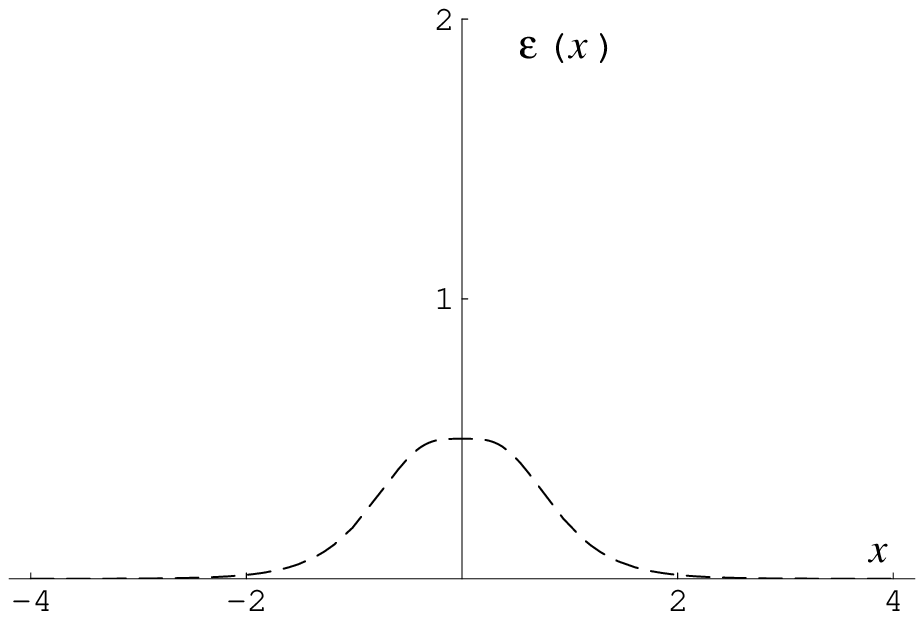, height=2.5cm}
\epsfig{file=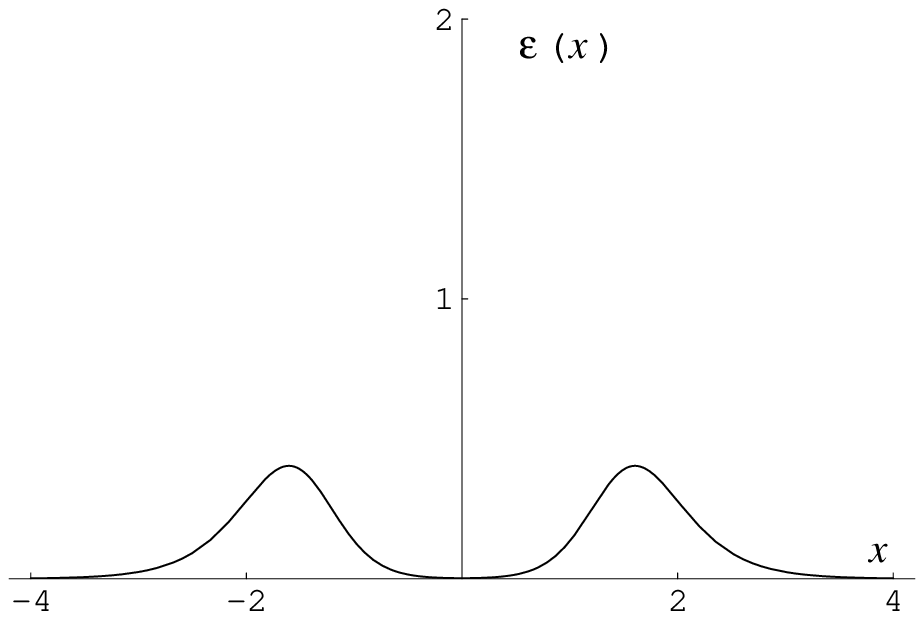, height=2.5cm} } \caption{\small
Energy density ${\cal E}^K[x;0,b]$ for {\it (a)} $b=0$, {\it (b)}
$b=\sqrt{0.5}$, {\it (c)} $b=1$ and {\it (d)} $b=\sqrt{30}$.}
\end{figure}

This qualitative description can be precisely established in an
analytic fashion by looking at the maxima of the energy density
${\cal E}^K[x;a,b]$. These can be found through a classical
analysis, applying the Cardano and Vieta formulas and Rolle's
theorem. We obtain the following conclusions:

\noindent 1. If $b^2\in [0,1]$, $x=0$ is the only critical point
(maximum) of ${\cal E}^K$ and ${\cal E}^K[0;0,b]={2\over
(b^2+1)^2}$. Therefore, $b^2$ measures the height of ${\cal E}^K$
in this regime where the two lumps are aggregate.

\noindent 2. If $b^2\in (1,\infty)$,  $x=0$ is a minimum. Because
${\partial{\cal E}^K\over\partial x}[x;0,b]={2{\rm
sinh}x\over[b^2+{\rm cosh}x]^5}P_3[{\rm cosh}x]$, where $P_3[{\rm
cosh}x]$ is a third order polynomial with real roots
$\pm\sqrt{r(b^2)}$, see \cite{Aai1}, we identify $x=\pm
m(b^2)=\pm{\rm arccosh}[1+r(b^2)]$ as the two maxima of ${\cal
E}^K$ obeying to the peak of the energy density of the two lumps.
$b^2$ measures (in a highly non-linear scale set by the known
function $r(b^2)$) the distance between lumps.

\section{Low-energy classical dynamics of BPS domain walls}
In this section, we recover the (3+1)-dimensional point of view
where our kinks become domain walls. The aim is to study the
low-energy classical dynamics of these BPS topological walls,
which can be understood as composed of the two basic link walls.
We shall focus on the $\sigma=\frac{1}{2}$ case, for which
analytical formulas are available.
\subsection{Adiabatic motion orthogonal to the wall}
We first analyze the motion orthogonal to the wall. In the case of
walls grown from kinks of a single real scalar field, this
analysis is not necessary because Lorentz invariance takes care of
the matter. Besides the $a$ coordinate, describing the motion of
the wall center of mass in the orthogonal direction ruled by
Lorentz symmetry, there is another parameter in the moduli space
of domain walls: the relative coordinate $b$. The dynamics of the
motion on the $b$-coordinate along the $x$-axis is non-trivial;
the dependence of $b$ on time precisely characterizes how the two
basic walls intersect and split on their way along the $x$-axis.

Starting from the Hamiltonian of the reduced system,
\[
H[\phi_1,\phi_2,\dot{\phi_1},\dot{\phi_2}]=\int \, dx {1\over
2}\left({\partial\phi_1^K\over\partial
x^0}{\partial\phi_1^K\over\partial
x^0}+{\partial\phi_2^K\over\partial
x^0}{\partial\phi_2^K\over\partial x^0}\right)+\int \, dx\,  {\cal
E}\qquad ,
\]
we apply the adiabatic hypothesis of Manton \cite{Manton} to study
the low-energy dynamics of topological defects as geodesic motion
in the moduli space. The smooth evolution on the moduli hypothesis
\[
\phi_1^K(x^0,x)=\phi_1^K[x;a(t),b(t)] \qquad , \qquad
\phi_2^K(x^0,x)=\phi_2^K[x;a(t),b(t)]
\]
is plugged into the action and, after integrating out the $x$
variable, we find that $S$ becomes the action for geodesic motion
in the kink moduli space with a metric inherited from the dynamics
of the zero modes: { \small\begin{eqnarray*} S=\int dt \left\{
\frac{1}{2}g_{aa}(a,b)\frac{da}{dt}\frac{da}{dt}+
g_{ab}(a,b)\frac{da}{dt}\frac{db}{dt}+
\frac{1}{2}g_{bb}(a,b)\frac{db}{dt}\frac{db}{dt}\right \} .
\end{eqnarray*}}
The components of the metric tensor are:
$g_{aa}(a,b)=\frac{1}{3},\  g_{ab}(a,b)=0,\
g_{bb}(a,b)=\frac{1}{3}h(b)$, where
\begin{equation}
h(b)=\frac{1}{4(b^4-1)^2}\left[ 2 b^6-5b^2+3\frac{\arctan \left(
\frac{\sqrt{1-b^4}}{b^2}\right)}{\sqrt{1-b^4}}\right]
\label{eq:metric}\quad .
\end{equation}
As expected, the metric is independent of the center of mass $a$.
Despite appearances, the behaviour of the metric is regular in the
transition of $b^2$ from lower to higher values than 1. For a
metric of the form given, the geodesics are easily found: they
are merely straight lines on the $a-{\bar b}$ plane:
\begin{eqnarray} a(t)=k_1t+k_2 \qquad , \qquad {\bar b}(t)=
k_1^\prime t+k_2^\prime=\int \, db\sqrt{h(b)} \label{eq:str}\qquad
. \end{eqnarray} $k_1$, $k_2$, $k_1^\prime$ and $k_2^\prime$ are
integration constants. It is worthwhile to use (\ref{eq:str}) to
express the geodesic orbits in the kink space:
\begin{equation}
{\bar b}=\int \, db\sqrt{h(b)}=\kappa_1a+\kappa_2 \qquad , \quad
\kappa_1={k_1^\prime\over k_1} \quad , \quad
\kappa_2=k_2^\prime-\kappa_1k_2 \label{eq:geod}
\end{equation}
There are two main types:

\noindent $\bullet$ Choosing $\kappa_1=0\equiv b={\rm constant}$
in (\ref{eq:geod}) we obtain geodesics describing free motion of
the center of mass without any variation in separation of the two
lumps, see Figure 4.

\noindent
\begin{figure}[htbp]
\centerline{ \epsfig{file=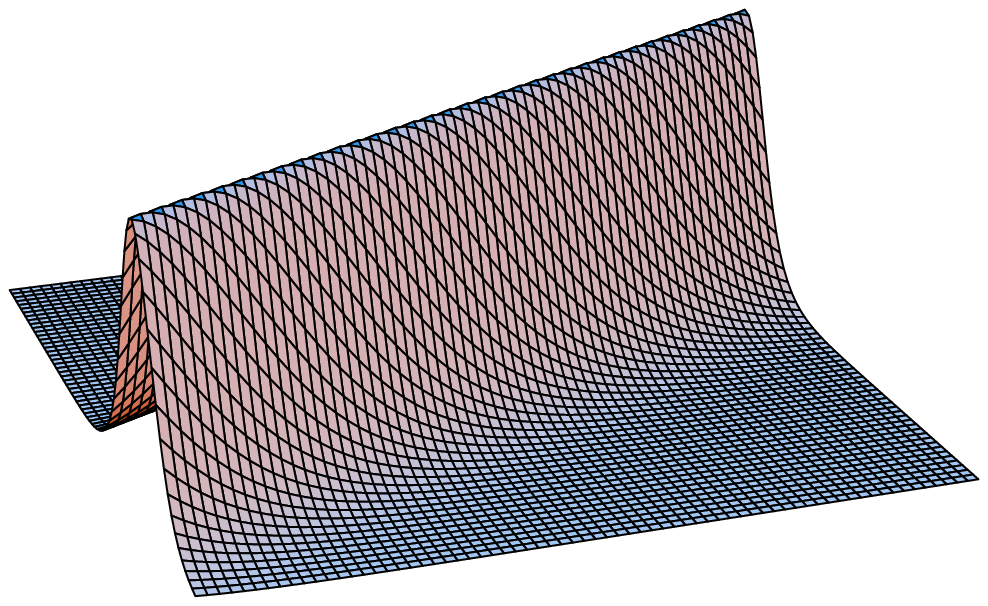, height=3cm}
\epsfig{file=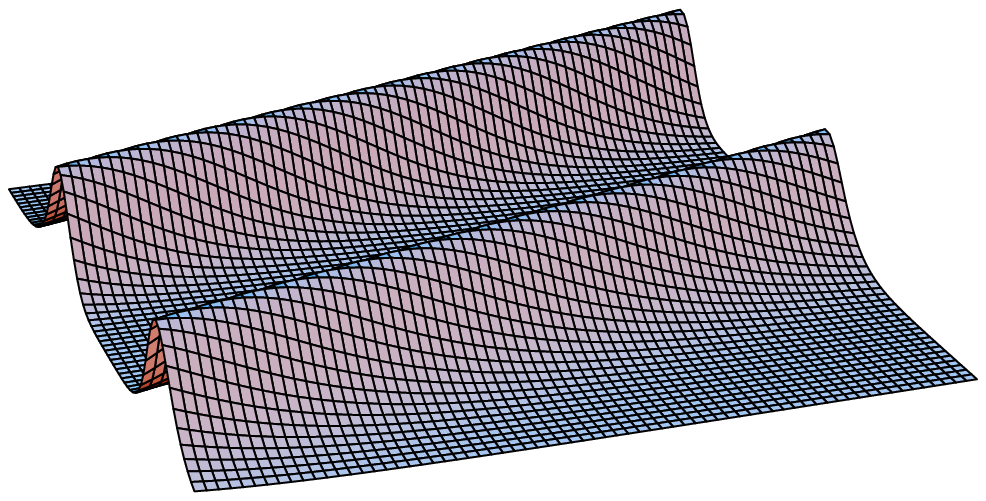, height=3cm} }\caption{\small Energy
density evolution along straight geodesic lines with
$b$=constant: {\it (a)} $b=0.9$, a single lump is moving {\it
(b)} $b=10$, synchronous motion of two lumps. Time runs from left
to right.}
\end{figure}

\noindent $\bullet$ If we choose $\kappa_1\neq 0$ the geodesics
also describe a non-trivial motion of the relative coordinate. A
MATHEMATICA numerical plot choosing $\kappa_1=3$, $\kappa_1=2$,
$\kappa_1=1$ whereas $\kappa_2$ is fixed by setting $b=0.1$ at
$t=-{k_2\over k_1}$ is shown in Figure 5(a). Clearly, these
geodesics describe exact solutions at the adiabatic limit for
intersecting walls. There is analogy with the scattering of
solitons in the sine-Gordon model, although, in this case,
shape-preserving collisions only occur in the topological sector
with a loop kink. As compared with similar phenomena, we find
hybrid behaviour in our system between the sine-Gordon and
$\lambda(\phi^4)_2$ models.

\begin{figure}[htbp]
\centerline{ \epsfig{file=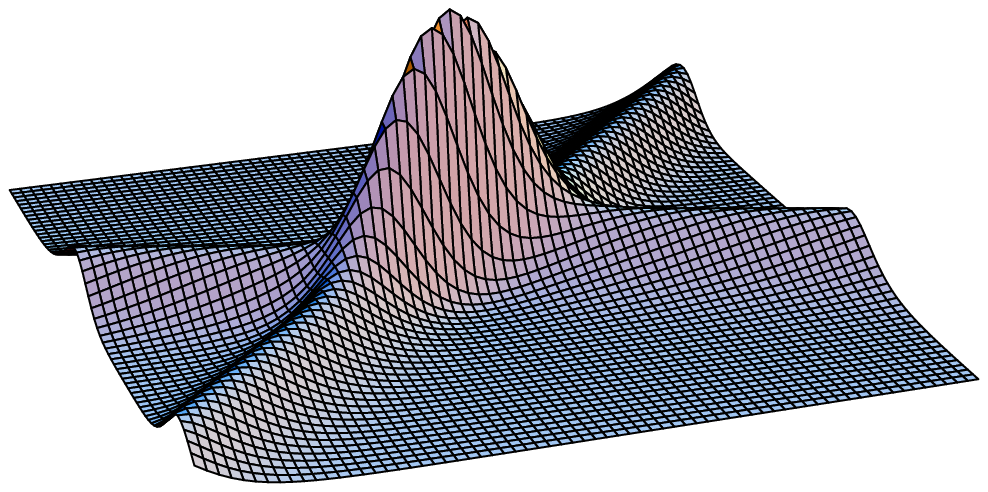,
height=3.5cm}\hspace{1cm}\epsfig{file=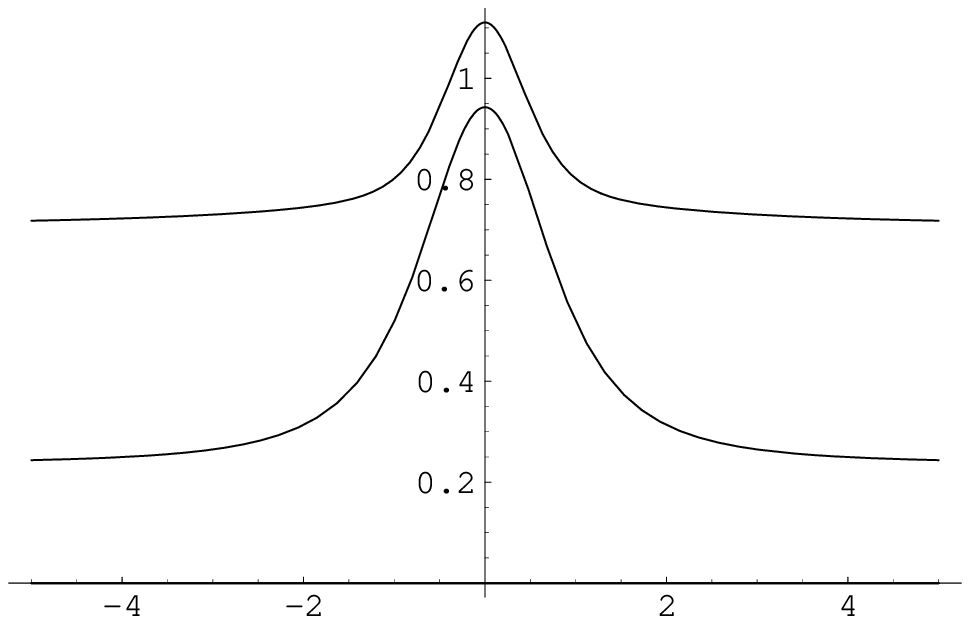,height=4cm}}\caption{\small
(a) Evolution of energy density along a generic geodesic curve.
(b) Plot of the functions $f^2(b)$ (up) and $f^1(b)$ (down).}
\end{figure}

\subsection{Effective action for intersecting walls}
The effective action for domain walls modeled on kinks without
internal structure is derived by expanding the action around the
classical solution and taking into account only the zero mode in
the direction orthogonal to the wall, see \cite{Gomis}. We proceed
along the same way to unveil the effective action induced by the
zero modes for intersecting walls. There are two zero modes in the
direction orthogonal to these composite two-branes. The collective
coordinates corresponding to these zero modes are precisely the
coordinates of the kink moduli space.

The Hessian driving the small fluctuations orthogonal to the
domain wall is
\begin{equation}
{\cal K}=\left( \begin{array}{cc}
\frac{d^2}{dx^2}-\frac{3b^2}{\cosh x+b^2}-\frac{6\sinh^2x}{(\cosh
x+b^2)^2}+2 & \frac{-6b\sinh x}{(\cosh x+b^2)^{\frac{3}{2}}}
\\\frac{ -6b\sinh x}{(\cosh x+b^2)^{\frac{3}{2}}} &
\frac{d^2}{dx^2}-\frac{3b^2}{2(\cosh x+b^2)}-
\frac{3\sinh^2x}{4(\cosh x+b^2)^2}+\frac{1}{2}\end{array} \right)
\label{eq:fhess}
\end{equation}
and by expanding the (3+1)-dimensional action restricted to the
Bose sector around the kink solutions
\[
\phi_1(\vec{x},t)=\phi_1^{TK2}[x;a,b]+\delta\phi_1(\vec{x},t)
\qquad , \qquad
\phi_2(\vec{x},t)=\phi_2^{TK2}[x;a,b]+\delta\phi_2(\vec{x},t)
\]
up to second order in small fluctuations we obtain:
\begin{displaymath}
S=\frac{1}{2}\int \, dtd^3\vec{x}\left\{  \delta\phi_a {\cal
K}_{ab}
\delta\phi_b+(\partial_t\delta\phi_1)^2-(\partial_y\delta\phi_1)^2-(\partial_z\delta\phi_1)^2
+(\partial_t\delta\phi_2)^2-(\partial_y\delta\phi_2)^2-(\partial_z\delta\phi_2)^2\right\}\
.
\end{displaymath}
Note that the metric found in the previous subsection implies that
the (non-dimensional) constant energy per unit of area, the
surface tension \footnote{Using full dimensional variables, where
the superpotential is $\bar{W}(\vec{\chi})=\lambda \left(
\frac{1}{3} \chi_1^3 -a^2 \chi_1 \right)+\frac{1}{2} \mu \chi_1
\chi_2^2$, we would have obtained $T={4\over 3}a^3\lambda$, see
\cite{Aai5}.} of the wall, is $T=\frac{1}{3}$.

Because the solutions only depend on $x$, we now attempt to
separate variables
\[
\delta\phi_1(x,y,z,t)=Z_1(x)X_1(y,z,t) \qquad , \qquad
\delta\phi_2(x,y,z,t)=Z_2(x)X_2(y,z,t)\qquad ,
\]
and because the spectrum of the (1+1)-dimensional Hessian has a
mass gap $\frac{a^2 \lambda^2}{4}$, at low energies the only
contribution to $Z$ comes from the zero modes. Therefore, the
effective action is
\begin{equation}
S_{eff}=T\int dtdydz  \left\{1+\sum_{j=1}^2{f^j(b)\over
T}\left[(\partial_tX_j)^2-(\partial_yX_j)^2-(\partial_zX_j)^2\right]\right
\}\qquad , \label{eq:eff}
\end{equation}
where the functions $f^1(b)$ and $f^2(b)$ are defined from the
zero modes:
\[
f^1(b)=\int dx \left({\partial\phi_1^{TK2}\over\partial
a}{\partial\phi_1^{TK2}\over\partial
a}+{\partial\phi_1^{TK2}\over\partial
b}{\partial\phi_1^{TK2}\over\partial b}\right)\quad , \quad
f^2(b)=\int dx \left({\partial\phi_2^{TK2}\over\partial
a}{\partial\phi_2^{TK2}\over\partial
a}+{\partial\phi_2^{TK2}\over\partial
b}{\partial\phi_2^{TK2}\over\partial b}\right) \quad .
\]
As in the case of the metric tensor components these integrals are
independent of the center of mass $a$ and can be performed by
changing variables to $u=e^{(x+a)}$:
\[
{f^1(b)\over T}={1\over
8}\left(\frac{2b^8+b^6-10b^4+2b^2+8}{(b^4-1)^2}+3
\frac{b^2(2b^4-b^2-2)}{(1-b^4)^{{5\over 2}}}{\rm arc
cot}(\frac{b^2}{\sqrt{1-b^4}})\right)
\]
\[
{f^2(b)\over T}={3\over
8}\left(\frac{b^2(2b^6+b^4-2b^2-4)}{(b^4-1)^2}-
\frac{2b^6-b^4-2b^2-2}{(1-b^4)^{{5\over 2}}}{\rm arc
cot}(\frac{b^2}{\sqrt{1-b^4}})\right)
\]
Again we obtain a regular answer: see the graphics of $f^1(b)$ and
$f^2(b)$ in Figure 5(b). Formula (\ref{eq:eff}) tells us that the
two Goldstone bosons $X_1(t,y,z)$ and $X_2(t,y,z)$ living inside
the wall feel a different tension that are functions of the
relative coordinate. The dependence of the surface tensions on how
far or how close the two basic lumps are follows the graphics in
Figure 5(b).

\section{One-loop renormalization of the surface tension: induced
repulsive forces}

Do quantum effects modify the picture that we have described? Is
the dynamics of domain walls in the quantum world different? To
answer these related questions, we develop a semiclassical
analysis of the domain walls in the Bose sector of the generalized
Wess-Zumino model.
\subsection{TK1 kink mass in the generalized Wess-Zumino model}
We start with the one-component topological kink arising when
$c=-\infty$. The second-order fluctuation operator around the TK1
kink is a \lq\lq diagonal" matrix-valued Schr\"{o}dinger operator:
\begin{equation} {\cal K}=\left( \begin{array}{cc}
-\frac{d^2}{dx^2}+4 -6 {\rm sech} ^2 x & 0 \\ 0 &
-\frac{d^2}{dx^2}+\sigma^2 -\sigma (\sigma+1){\rm sech} ^2 x
\end{array} \right)\qquad .
\end{equation}
There are contributions of the \lq\lq tangent" and \lq\lq
orthogonal" fluctuations to the semi-classical kink mass:
\[
\Delta M({\rm TK1})=\Delta M({\cal K}_{11})+\Delta M({\cal
K}_{22}) \qquad .
\]

\subsubsection{One-loop correction to the TK1 kink mass}

We shall apply the generalized DHN formula
\begin{eqnarray}
\Delta M({\cal K}_{aa})&=&\frac{\hbar m}{2} \left[
\sum_{i=0}^{l-1} \omega_i+s_{l}\omega_{l}-\frac{v_a}{2}  +
\frac{1}{\pi} \int_0^\infty \!\!\! dq \frac{\partial
\delta_a(q)}{\partial q} \sqrt{q^2+v_a^2}- \frac{\left< V_{aa}(x)
\right> }{2 \pi} \right]+ \nonumber \\&+& \hbar m\frac{\left<
V_{aa}(x) \right> }{8 \pi} \int_0^\infty \!\!
\frac{dk}{\sqrt{k^2+v_a^2}}
\end{eqnarray}
that was derived in \cite{Aai5}. The following conventions are
defined:
\[
\displaystyle v_a^2=\left. \frac{\delta^2
U}{\delta\phi_a^2}\right|_{\vec{\phi}_{V^\pm_1}}
\hspace{0.5cm};\hspace{0.5cm} V_{aa}(x) =
v_a^2-\left.\frac{\delta^2
U}{\delta\phi_a^2}\right|_{\vec{\phi}_{TK1}}; \hspace{0.5cm}
\left<V_{aa}(x)\right>=\int_{-\infty}^\infty \!dx V_{aa}(x) \qquad
,
\]
which give $v_1^2=4$, $V_{11}(x)={6\over{\rm cosh}^2x}$, $
v^2_2=\sigma^2$ and $V_{22}(x)={\sigma(\sigma+1)\over{\rm
cosh}^2x}$ in the case of the TK1 kink of the generalized
Wess-Zumino model.

\noindent $\bullet$ For the tangent fluctuations, we recover the
old result of Dashen, Hasslacher and Neveu:
{\small
\begin{eqnarray*}
\Delta M({\cal K}_{11}^{\rm TK1}) &=& \hbar m\left(
\frac{\sqrt{3}}{2}-\frac{1}{2 \pi} \int_{-\infty}^\infty
\hspace{-0.3cm} dq \frac{3 \sqrt{q^2+4} (q^2+2)}{q^4+5 q^2+4}+
\frac{3 }{2 \pi} \int_{-\infty}^\infty \frac{d k}{\sqrt{k^2+4}} -
\frac{1}{4 \pi} \int_{-\infty}^\infty \, dx \, 6 \, {\rm sech} ^2
x\right)\\&=&\hbar m\left(\frac{1}{2\sqrt{3}}-\frac{3}{\pi}
\right)
\end{eqnarray*}
}

\noindent $\bullet$ The contribution of orthogonal fluctuations is
more difficult to compute. There are even and odd phase shifts,
\[
\delta_2^{\pm} (q)=\frac{1}{4}{\rm arctan}\left(\frac{{\rm
Im}(T(q)\pm R(q))}{{\rm Re}(T(q)\pm R(q))}\right) ,
\]
to be read from the transmission and reflection coefficients
\[
T(q)=\frac{\Gamma(\sigma+1-iq)\Gamma(-\sigma-iq)}{\Gamma(1-iq)\Gamma(-iq)}\quad
; \quad
R(q)=\frac{\Gamma(\sigma+1-iq)\Gamma(-\sigma-iq)\Gamma(iq)}
{\Gamma(1+\sigma)\Gamma(-\sigma)\Gamma(-iq)}\qquad .
\]
The spectrum of ${\cal K}_{22}$ ,
\[
{\rm Spec}\, ({\cal K}_{22})= \left\{
\begin{array}{lcl} \cup_{i=0,1,...,I[\sigma]}\{\omega_i= i(2\sigma-i) \} \cup
\{q^2+\sigma^2\}_{q\in {\Bbb R}^+}  && \mbox{if} \hspace{0.4cm}
\sigma \notin {\Bbb N} \\ \cup_{i=0,1,...,\sigma-1}\{\omega_i=
i(2\sigma-i) \} \cup \{\omega_{l=\sigma}=\sigma^2
\}_{s_{l=\sigma}=\frac{1}{2}} \cup \{q^2+\sigma^2\}_{q\in {\Bbb
R}^+}  && \mbox{if} \hspace{0.4cm} \sigma \in {\Bbb N}
\end{array} \right\} \qquad ,
\]
shows different patterns according to whether $\sigma$ is an
integer or not; in the first case the reflection coefficient is
zero and there is a half-bound state. In any case, one needs the
formula
\begin{eqnarray}
\frac{\partial \delta_2(q)}{\partial q}&=& -\frac{i}{2}\left[
e^{-i2\delta_2^+} \frac{\partial e^{i 2 \delta_2^+}}{\partial q}+
e^{-i 2 \delta_2^-} \frac{\partial e^{i 2 \delta_2^-}}{\partial
q}\right]= \nonumber \\ &=& 2 Re[\psi(iq)-\psi(-\sigma+i
q)]+\frac{\pi}{2 \sinh^2 \pi q \csc 2 \pi \sigma+\tan \pi \sigma}
 \label{eq:derfase}
\end{eqnarray}
to numerically compute: {\small\begin{eqnarray*}  \Delta M({\cal
K}_{22})
 &&= \frac{\hbar m}{2} \left[ \sum_{i=0}^{I[\sigma]}
\sqrt{i(2\sigma-i)} \!-\!\frac{\sigma}{2}\! +\! \frac{1}{\pi}
\int_0^\infty \hspace{-0.4cm} dq \left( \frac{\partial
\delta_2(q)}{\partial q} \sqrt{q^2\!+\!\sigma^2}
+\frac{\sigma(1+\sigma)}{\sqrt{q^2\!+\!\sigma^2}} \right) \!-\!
\frac{\sigma(\sigma+1) }{\pi} \right] \hspace{0.3cm} \mbox{if}
\hspace{0.3cm} \sigma \notin {\Bbb N} \nonumber \\ &&= \frac{\hbar
m}{2} \left[ \sum_{i=0}^{\sigma-1} \sqrt{i(2\sigma-i)} +\!
\frac{1}{\pi} \int_0^\infty \hspace{-0.4cm} dq \left(
\frac{\partial \delta_2(q)}{\partial q} \sqrt{q^2\!+\!\sigma^2}
+\frac{\sigma(1+\sigma)}{\sqrt{q^2\!+\!\sigma^2}} \right) \!-\!
\frac{\sigma(\sigma+1) }{\pi} \right] \hspace{0.3cm} \mbox{if}
\hspace{0.3cm} \sigma \in {\Bbb N} \qquad .
\end{eqnarray*}}
In Ref. \cite{Aai5}, a Table is offered with the result for
$\Delta M (TK1)$ and values of $\sigma $ between $0.4$ and $3.3$.

It is also possible to apply the formula
\begin{equation}
\Delta M ({\cal K}) =\hbar m[\Delta_0 +D_{n_0}] \left\{
\begin{array}{l} \Delta_0=-\displaystyle\frac{j}{2\sqrt{\pi}}  \\
D_{n_0}=-\displaystyle\sum_{a=1}^2\sum_{n=2}^{n_0-1}
\displaystyle\frac{[a_{n}]_{aa}({\cal K})}{8\pi}
\displaystyle\frac{\gamma[n-1,v_a^2]}{v_a^{2n-2}}\quad , \quad
n_0\in{\Bbb N}
\end{array}\right\}\qquad ,
\label{asymp}
\end{equation}
which was derived using zeta function regularization methods -as
those developed in \cite{Bordag} and applied to supersymmetric
kinks- in References \cite{Aai0} and \cite{Aai2}, to find $\Delta
M (TK1)$. Here, $j=2$ is the number of zero modes,
$\gamma[n-1,v_a^2]$ are incomplete Gamma functions, and
$[a_{n}]_{aa}({\cal K})$ are the Seeley coefficients of the
high-temperature expansion for the heat kernel of the ${\cal K}$
operator. Figure 6 (left) the good agreement between the exact and
the asymptotic result for $\sigma>1$.

\begin{figure}[htbp]
\centerline{\epsfig{file=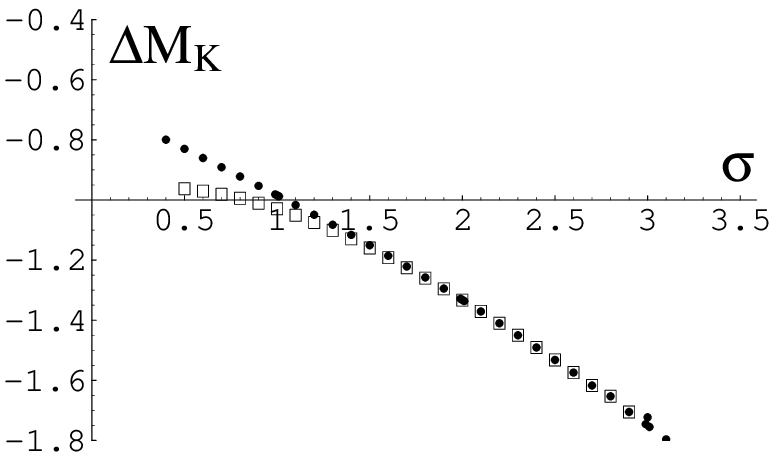,height=3.5cm}\hspace{1.2cm}
\epsfig{file=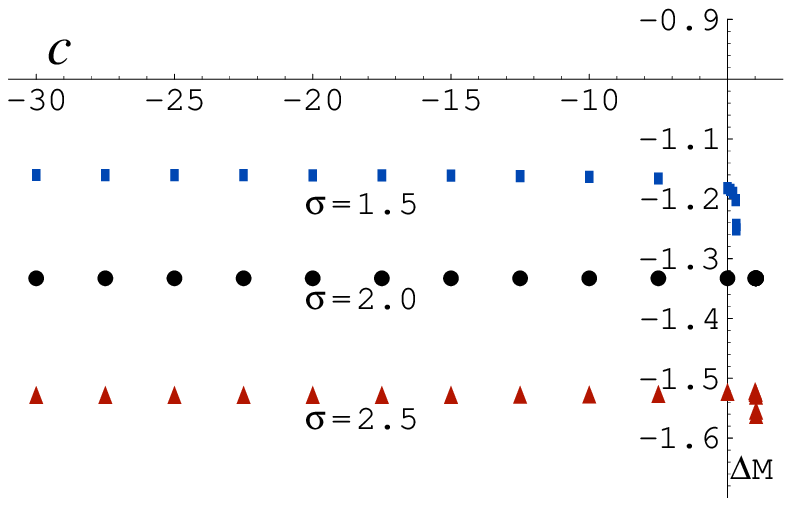,height=3.5cm}} \caption{\small {\it
One-loop correction to the one-component topological kink (TK1)
mass in units of $\hbar m$. $\bullet$,  DHN formula . $\Box$,
asymptotic series (left) The One-loop Quantum Mass Correction in
the cases $\sigma=1.5$, $\sigma=2.0$ and $\sigma=2.5$ (right)}}
\end{figure}

\subsection{Semi-classical masses of kink families}
We now try to compute the one-loop correction to the classical
mass for the whole kink family

This task is easy if $\sigma=2$. Although the family of
Schr\"{o}dinger operators governing the small fluctuations around
the TK2 kinks is non-diagonal,
\[
{\cal K}(b)= \left( \begin{array}{cc}
-\frac{d^2}{dx^2}+6\frac{\sinh^2(2x)+b^2-1}{(\cosh(2x)+b)^2}-2 &
12\sqrt{b^2-1}\frac{\sinh(2x)}{(\cosh(2x)+b)^2}
\\[0.2cm]
 12\sqrt{b^2-1}\frac{\sinh(2x)}{(\cosh(2x)+b)^2} &
-\frac{d^2}{dx^2}+6\frac{\sinh^2(2x)+b^2-1}{(\cosh(2x)+b)^2}-2\end{array}
\right) \qquad ,
\]
a rotation of $45^0$ in the internal space ${\Bbb R}^2$,
$\vec{e}_1=\frac{1}{\sqrt{2}}(\vec{\varepsilon}_1+\vec{\varepsilon}_2)$,
$\vec{e}_2=\frac{1}{\sqrt{2}}(\vec{\varepsilon}_1-\vec{\varepsilon}_2)$,
shows that the system is uncoupled. Writing
$\vec{\phi}=\psi_1\vec{\varepsilon}_1+\psi_2\vec{\varepsilon}_2$,
we have that:
\[
T_{\sigma=2}=\frac{1}{2}
\left(\frac{d\psi_1}{dx}\right)^2+\frac{1}{2}\left(\frac{d\psi_2}{dx}\right)^2
\qquad , \qquad
U_{\sigma=2}=4\left(\psi_1^2-\frac{1}{8}\right)+4\left(\psi_2^2-\frac{1}{8}\right)
\]
and the degenerate kink family is given as:
\[
\vec{\phi}_{TK2^{*}}[x;a_1,a_2]=\frac{(-1)^\alpha }{2\sqrt{2}}
\tanh(x+a_1)\vec{\varepsilon}_1+ \frac{(-1)^\beta}{2\sqrt{2}}
\tanh(x+a_2)\vec{\varepsilon}_2 \quad .
\]
The alternative form of the Hessian is:
\[
{\cal K}(a_1,a_2)= \left( \begin{array}{cc}
-\frac{d^2}{dx^2}+4-\frac{6}{\cosh^2(x+a_1)} & 0
\\ 0 &
-\frac{d^2}{dx^2}+4-\frac{6}{\cosh^2(x+a_2)}\end{array}
\right)\qquad .
\]
Therefore, $\Delta M(TK2^{*}[a_1,a_2])=\hbar
m(\frac{1}{\sqrt{3}}-\frac{6}{\pi})$. The kink degeneracy is not
broken by quantum fluctuations at the one-loop level.

For generic $\sigma$ there are no analytical solutions available.
We can however solve the first-order equations (\ref{eq:tra1}) by
standard numerical methods and setting, for example, the \lq\lq
initial" conditions:
\[
\phi_1(0)=0 \qquad , \qquad \frac{\sigma}{2(1-\sigma)}\phi_2^2(0)
-\frac{c}{2\sigma}|\phi_2(0)|^{\frac{2}{\sigma}}=\frac{1}{4}
\qquad .
\]
The polynomial kink solutions thus generated allow one to compute
the coefficients $[a_n]_{aa}({\cal K})$. The results obtained via
this numerical procedure are shown in Figure 6 (right). There is a
breaking of the degeneracy for values of $c$ close to $c^S$ if
$\sigma\neq 2$. The mass correction is lower when the two basic
lumps are far apart; henceforth, repulsive forces are induced by
the quantum fluctuations.

The $\sigma={1\over 2}$ case provides us with a qualitative
understanding of what is going on. The plot of the diagonal
components of the potential in the Schr\"{o}dinger operator
(\ref{eq:fhess}) for several values of $c$ shows that the
potential in the second component starts to be repulsive at the
value of $c$ where the two lumps start to split.
\begin{figure}[htbp] \centerline{
\epsfig{file=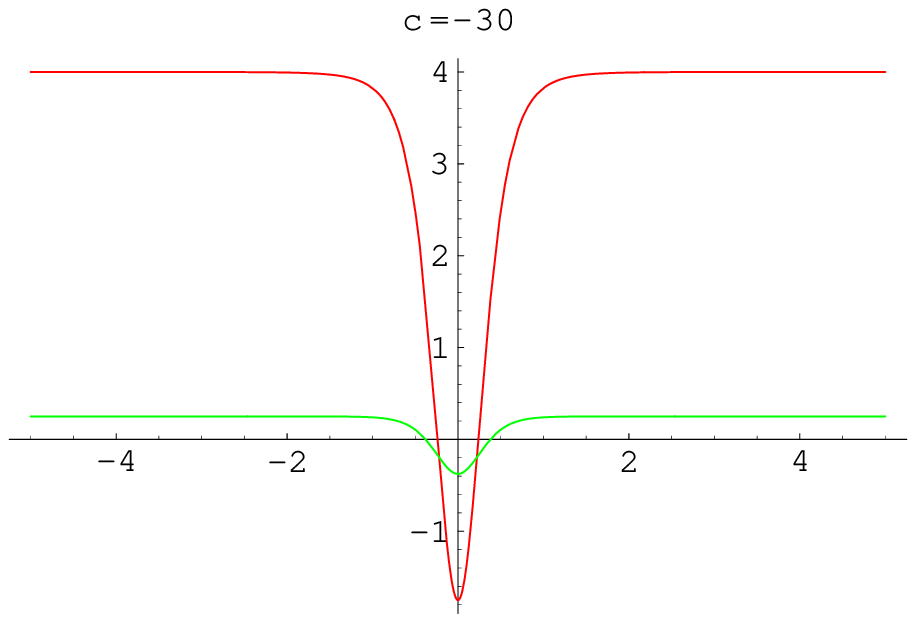,height=2.5cm}\quad
\epsfig{file=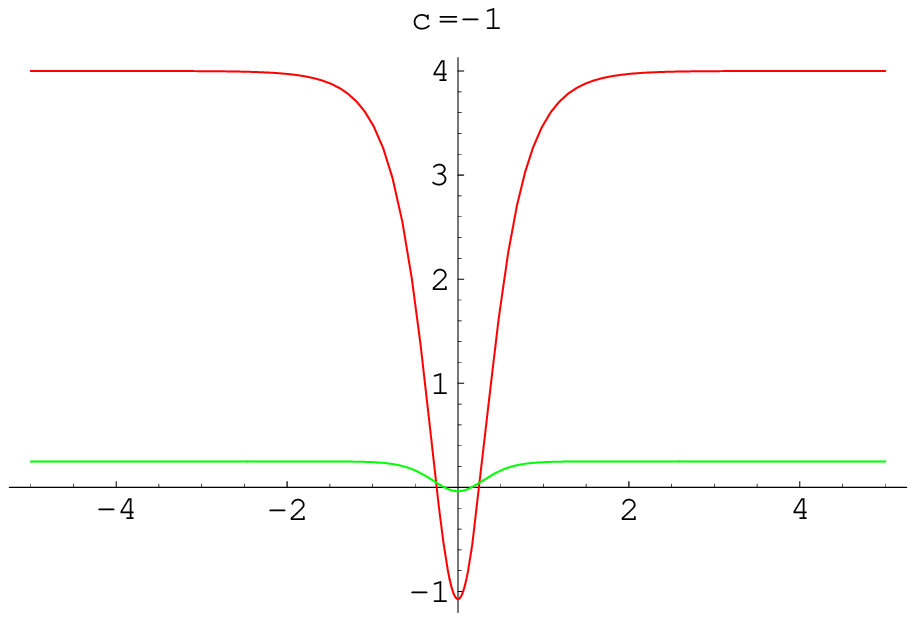,height=2.5cm}}
\centerline{\epsfig{file=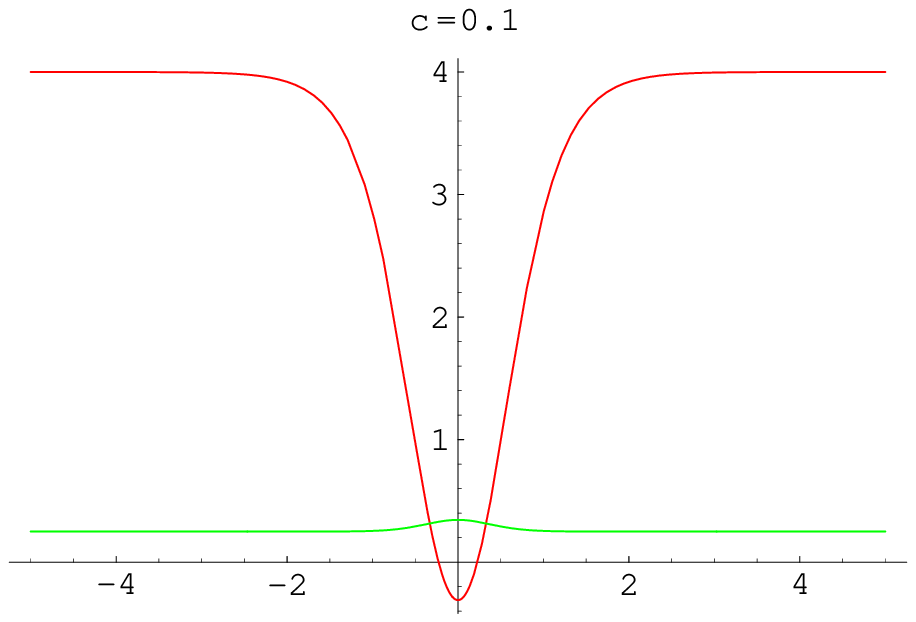,height=2.5cm}\quad
\epsfig{file=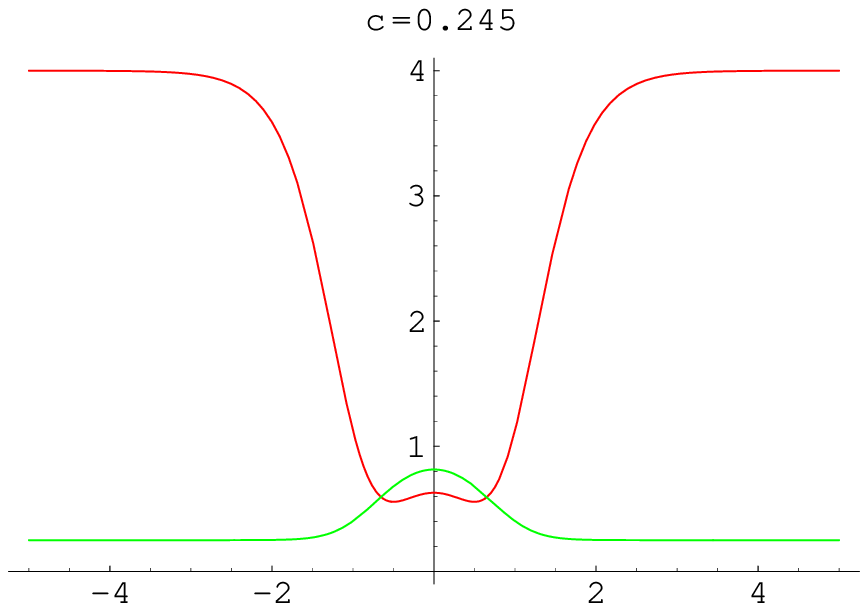,height=2.5cm}\quad
\epsfig{file=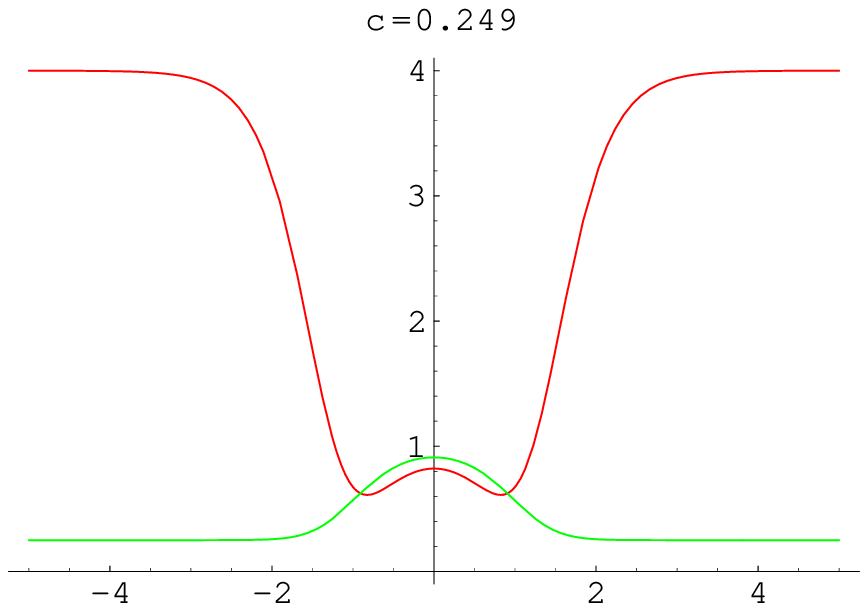,height=2.5cm}} \caption{\small \it
Diagonal components of the potential for c=-30, c=-1, c=0.1,
c=0.245 and c=0.249.}
\end{figure}

\section*{Acknowledgements}
We warmly thank J. Gomis for prompting us to think about the
effective action developed by our intersecting walls. We are also
grateful to him for sending us his unpublished Leuven Lectures on
Branes.

\end{document}